\documentstyle[bo99,epsfig]{article}

\def\Ms{M_\odot}

\def\rs{\rm s}
\def\rs1{\rm s^{-1}}

\def\rcm{\rm cm}
\def\rcm2{\rm cm^{-2}}
\def\deg{\rm ^{\circ}}

\def\c2r{\chi^2_\nu}
\def\simlt{$\; \buildrel < \over \sim \;$}
\def\ltsima{\lower.5ex\hbox{\simlt}}
\def\simgt{$\; \buildrel > \over \sim \;$}

\def\gtsima{\lower.5ex\hbox{\simgt}}
\def\etal{{\it et al. }}
\def\apj{Ap.J.}

\title{X-ray Afterglows of Gamma-Ray Bursts}

\author{Luigi Piro}

\affil{
Istituto Astrofisica Spaziale, C.N.R., Via Fosso del Cavaliere,
00133 Roma, Italy}

\begin{document}

\maketitle

\begin{abstract}

The afterglow emission has become the main stream of Gamma-Ray
 burst research since its discovery three years ago. With the distance-scale 
enigma solved, the study of the late-time
GRB emission is now the most promising approach 
to disclose the origin of these explosions
     and their relationship with the environment of the host 
galaxy in the early phase of the Universe. In this 
contribution I will review X-ray observations and their implication on our
     undertstanding on the GRB phenomenon. 
These measurements are providing
a direct probe into the nature of the progenitor and
a measurement of the GRB beaming properties, crucial to 
establish the total energy output. 
Some evidence of iron lines connects the GRB explosion with
massive progenitors, thence with star-forming regions.
Furthermore a comparison of
the spectral  properties with the temporal evolution 
indicates that the fireball 
expansion should not be - on average - highly collimated,
with a jet angle $>10 \deg$. 


\end{abstract}

\section{Introduction}
Gamma-Ray Bursts (GRB) were discovered in 1969 (Klebesadel \etal 1973)
by the Vela satellites, deployed by USA to verify the 
compliance of USSR to the nuclear test ban treaty. 
In the following 28 years thousands of events have been observed by several
satellites, leading to a good characterization of the 
global properties of this phenomenon.  A big step in this area was
achieved with BATSE (Fishman \etal 1994)
The isotropical distribution of the events in the sky 
(Fishman \& Meegan 1995) was suggestive
of an extragalactic origin, but a direct measurement of the distance in a 
single object was not available.
What was lacking was a {\it fast AND precise}
position, where the {\it Holy Grail} of 
GRB scientists, i.e. the {\it counterpart}, could have
been searched for at all wavelenghts with more chances
to catch it. 
This  was achieved  in 1996, with  observations
of GRB by BeppoSAX.

\section{Gamma-ray bursts in the Afterglow Era}

\subsection{The first afterglow: GB970228}
Before BeppoSAX (Piro \etal 1995, Boella \etal 1999) 
GRB astronomy has proceeded on
a statistical approach and the only information
gathered  was limited to the tens of seconds
of the GRB: the subsequent evolution was
completely unknown.
The operations for a prompt follow up of GRBs became operative
on December 1996, after  an off-line analysis of a GRB
(GRB960720: Piro \etal 1998a) 
had demonstrated the designed capability of the mission.  
The first opportunity was on January 11, 1997: GRB970111. 
The field was pointed with the NFI 16 hours after the GRB.
The possible association of one of the
faint sources
found in the error box with the GRB was under scrutiny
(Feroci \etal 1998), when 
on February 28, 1997, another event, GRB970228, was detected
by BeppoSAX GRBM and WFC. 
The NFI were pointed to the GRB location  
 8 hours after burst. 
A previously unknown X-ray source
was detected in the field of view of the LECS and MECS instruments
with a flux in the 2-10 keV energy range  
of  $3\times 10^{-12}~ erg~ cm^{-2}~ s^{-1}$.
The new source appeared to be fading away during the observation.
On March 3 we performed another observation that confirmed that
the source was quickly decaying : at that time its flux was a factor of
about 20 lower than the first observation 
 (Figure 1). 
This was the first detection of an "afterglow" of a
GRB (Costa \etal 1997).

\begin{figure}
\centerline{\epsfig{file=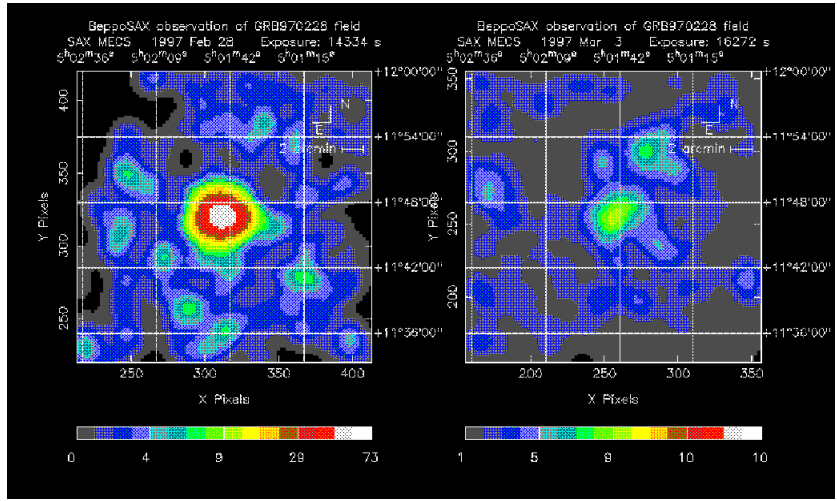, angle=-90,
width=11cm}}
\caption{BeppoSAX MECS images of the GRB970228 afterglow, 8 hours after the
GRB (left) and 3 days after the GRB (right) (from Costa \etal 1998)}
\label{grb970228_img}
\end{figure}
           
The flux of the source appeared to decrease following a power
law dependence on time ($\sim t^{-\alpha}$) with index 
$\alpha=(1.3\pm0.1)$. Further X-ray observation with the
X-ray satellites ASCA and ROSAT  detected the source about one week later 
with a flux consistent with the same law 
(Yoshida \etal 1997, Frontera \etal 1998a). 
This kind of temporal 
behaviour agrees with the general predictions of the fireball models
for GRBs (e.g.M\'esz\'aros P. \& Rees 1997, Vietri 2000 ).
A backward extrapolation of this power law decay 
(Fig. 2) is
consistent with the X-ray flux measured during the burst, suggesting
that the afterglow started soon after the GRB.
Another important result came from the spectral analysis of 
the X-ray  afterglow. It excluded a black body   
emission, therefore arguing against a
model in which the radiation comes from the cooling of the
surface of a neutron star
(Frontera \etal 1998b).

\begin{figure}
\centerline{\epsfig{file=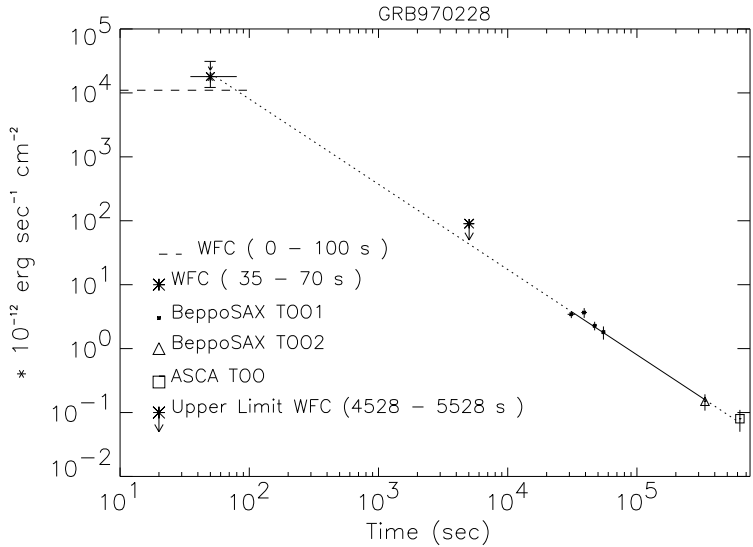,
width=10cm}}
\caption{BeppoSAX decay curve of the GRB970228 afterglow in the 2-10 keV
energy range, obtained with
the WFC and the NFI. The result of the ASCA observation is also shown
at the bottom right(from Costa \etal 1998).}
\label{grb970228_decay}
\end{figure}

While the X-ray monitoring of GRB970228 was going on, an observational
campaign of the same object was simultaneously 
started with the most important 
optical telescopes.
This campaign led to the discovery (van Paradijs \etal 1997) of an 
optical transient 
associated with the X-ray afterglow. 
As in the X-ray domain, the
optical flux of the source showed a decrease well described by a 
power law with index -1.12 (e.g. Garcia \etal 1998), again in agreement
with the general predictions of the fireball model.
The images taken with the Hubble Space Telescope (HST) 
(Fruchter \etal 1997, Sahu \etal 1997)
showed the presence of a nebulosity 
around the optical transient. However the nebulosity was very weak,
and it was not possible to disentangle whether
 it was associated with the host galaxy
(extragalactic origin)
or with a transient diffuse emission representing the residual of the
explosion (galactic origin).

\begin{figure}
\vspace{2cm}
\centerline{\epsfig{file=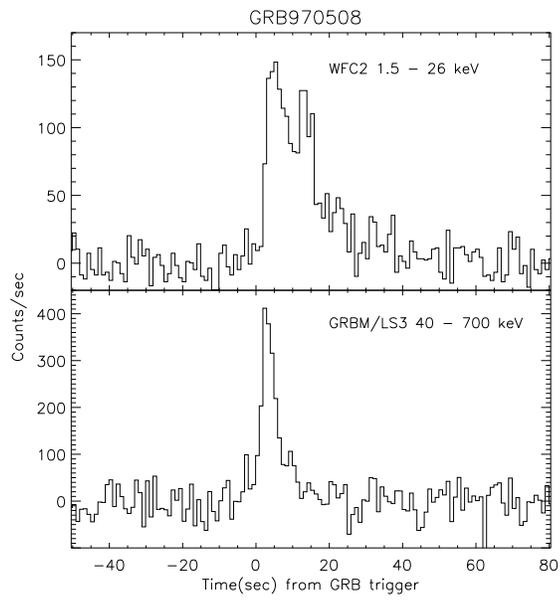,
width=8.cm}}
\vspace{-1cm}
\caption{WFC and GRBM light curves of GRB970508 (from Piro \etal 1998b)}
\label{grb970508_lc}
\end{figure}

\subsection{The first measurement of  redshift: GRB970508}


On 8 May 1997 the second breakthrough arrived with
GRB970508 (Piro \etal 1998b), detected just few minutes 
before the satellite was passing over
the ground station in Malindi. This opportunity
and the experience gained
from previous events allowed to
point the BeppoSAX NFI on source 
5.7 hours after the burst, while  optical observations
started 4 hours after the burst.

The early detection of the optical
 transient
(Bond 1997),
and  its relatively bright magnitude allowed a spectroscopical
measurement of its optical spectrum with the Keck telescope
(Metzger \etal 1997).
 The spectrum revealed the
presence of FeII and MgII absorption lines  at a
redshift of  $z=0.835$, attributed to the presence of a
galaxy between us and the GRB, and therefore demonstrated
that GRB970508 was at a cosmological distance.

\begin{figure}
\vspace{-3cm}
\centerline{
\epsfig{file=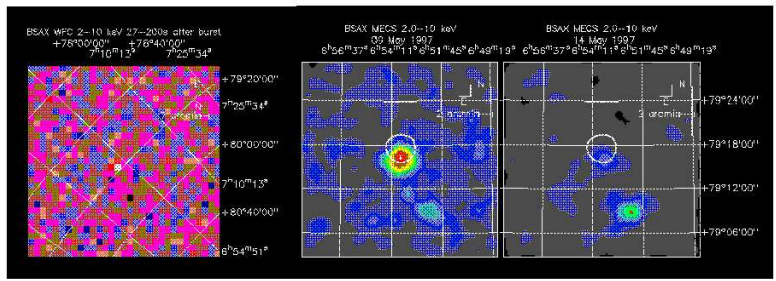,width=11cm,
angle=-270}
}
\vspace{-1.5cm}
\caption{The X-ray afterglow of GB970508 (from Piro \etal 1998b):
Time sequence of images of the field of GB970508 observed by the 
WFC2 (left image, 27--200 s after the burst), MECS(2+3)
on May 9 (center image, 6 hours after the GRB),  and 
MECS(2+3) on May
14 ( after 6 days). 
The WFC2 show the presence of the afterglow that was then
detected by the LECS and MECS 
(1SAXJ0653.8+7916 visible in the 99\% error circle of the WFC).
Note the decrease in intensity 
between the two MECS observations, as compared
to 1RXSJ0653.8+7916,  the source in the lower right corner
}
\label{bigfig}
\end{figure}

\subsection{GRB are long-lasting phenomena ... after all! }
The BeppoSAX observation of GB970508 has also changed
our view of the GRB {\it phenomenon}. 
The old concept of a brief - sudden release
of luminosity concentrated in few seconds does not stand  
the new information provided by BeppoSAX.
Indeed the name {\it afterglow} attributed to the X-ray  
emission observed after the event is somewhat misleading.
This is clear when one considers the energy produced in
the afterglow phase, which turns out to be comparable to
that of the GRB.
In order to compute the energy emitted in the afterglow phase
it is necessary to integrate in time its luminosity and
it is then crucial to know {\it when} the afterglow starts.
A detailed analysis of the data of the WFC of GB970508
(Piro \etal 1998b)  shows that the
X-ray emission  is  present (Fig. 4)
even when the signal of the light curve disappears
in to the noise at $\sim 30$ s (Fig.3) and remains visible
for at least 2000 seconds, when the
flux goes below the sensitivity of the WFC (Fig. 5). 
The conclusion is then that the afterglow phase starts immediately
after (or even before) than the prompt emission settles down. 
The energy emitted in the afterglow in
X-rays turns out to be a substantial fraction ($\sim 40-50\%$) of the
energy produced by the GRB.
Furthermore the light curve in Fig. 5 shows a rebursting event starting 1 day 
after the initial burst, an evidence that indicates that the source
of the energy can re-ignite on long time scales (Piro \etal 1998b).

\begin{figure}
\centerline{\epsfig{file=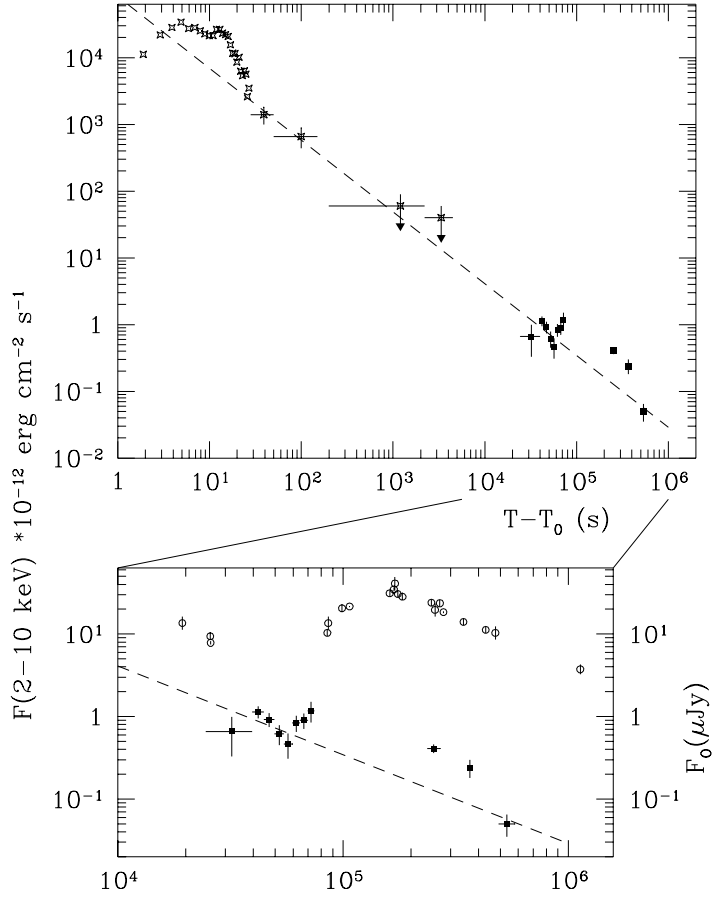,
width=12cm}}
\caption{Top panel: Decay law of the GRB970508 afterglow as
detected by the BeppoSAX
WFC and NFI. The WFC provided the data up to 5.000 seconds after the burst. 
Bottom panel: enlargement of the X-ray decay law and comparison
with the simultaneous time history of the optical transient (open circles)
(from Piro \etal 1998b)}
\label{grb970508_decay}
\end{figure}

\section{The largest and most distant explosions
since the Big Bang}

The GRB observed by BeppoSAX on December 14, 1997 if, on one hand,
has consolidated the extragalactic origin of GRB, on the other has
underlined the problem of the energy budget and, ultimately, 
of the nature of the ``central engine''.

The chain of steps leading to the identification of the counterpart of
GRB971214 (Dal Fiume \etal 1999) and its distance was the same of previous BeppoSAX observations.
With a redshift z=3.42 (Kulkarni \etal 1998a) this GRB and its host galaxy
are at a distance that corresponds to a look-back time of about
85\% of the present age of the Universe. At this distance, the luminosity
would be about $3\times 10^{53} erg\ s^{-1}$, were the emission isotropic.
This was the highest luminosity ever observed from any celestial
source.  Initially the huge luminosity appeared to be not 
compatible with the energy available in the coalescence of 
neutron-star mergers (Kulkarni \etal 1998a), unless beaming were invoked. 
Other alternative energetic models are based on the  death
of extremely massive stars, leading to an explosion orders of magnitude more
energetic than a supernova, hence named {\it hypernova} (Paczynski 1998,
Woosley 1993).
However it was shown (M\'esz\'aros  \& Rees   1998a) 
that all these progenitors, whether
Neutron Star - Neutron Star mergers or hypernovae, eventually go through
the formation of a same Black Hole/torus system, from which the energy is 
extracted to form the GRB. The radiation physics and energy of all mergers
and hypernovae are then, to order of magnitude the same, and still compatible
with the luminosity observed in GRB971214.

\begin{figure}
\centerline{\epsfig{file=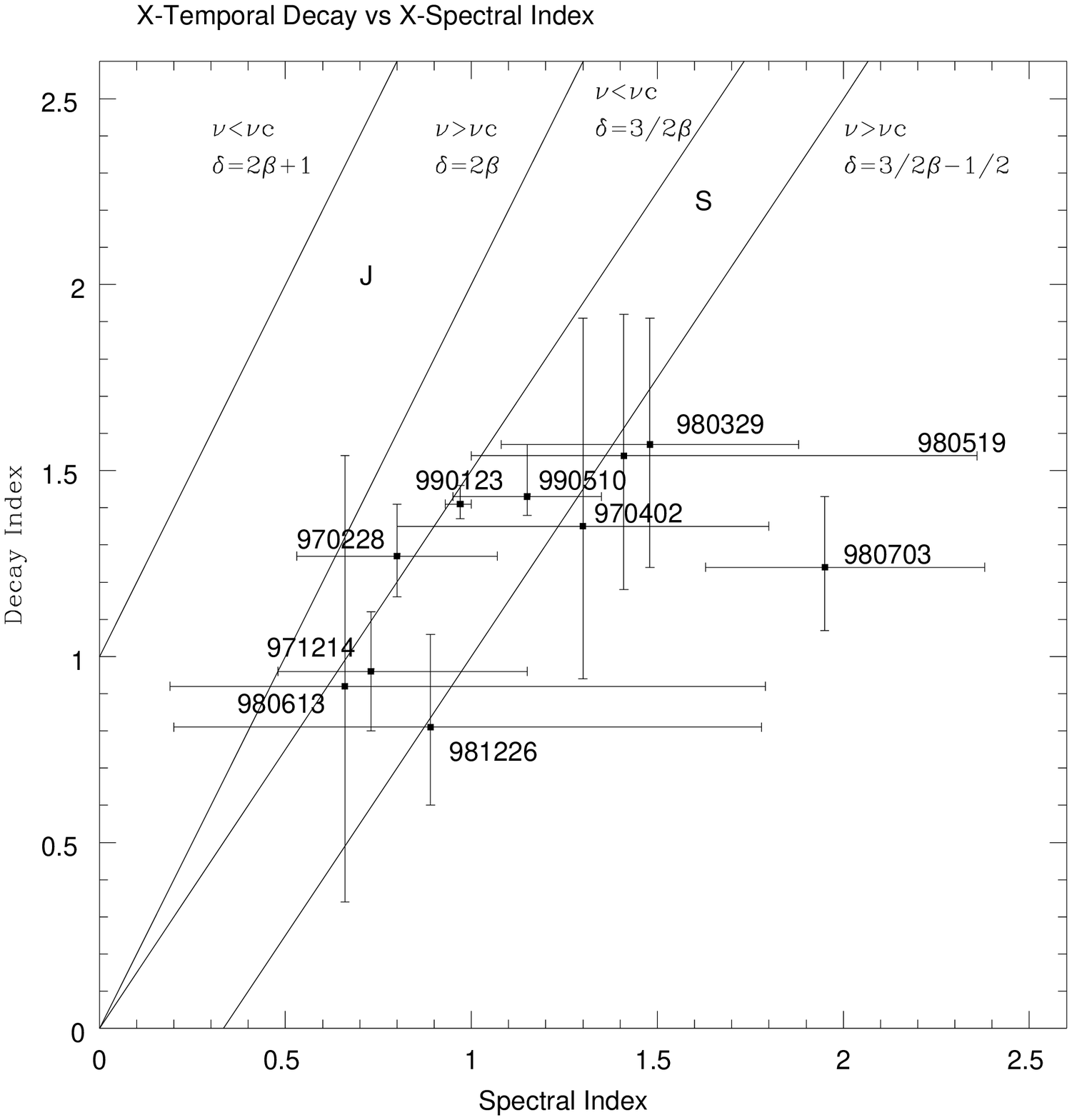,
width=9cm}}
\caption{
Decay vs spectral (photon index) power law slopes for
11 X-ray afterglows observed by BeppoSAX.
The continuous lines trace the expected relationship
in the cases of spherical (S)  and jet (J) expansion
in a uniform density medium}
\label{alfadeltax}
\end{figure}

\section{To beam or not to beam....}

The energy problem became much more severe with
 GRB990123, again one of the BeppoSAX
GRB (Heise \etal 1999, Piro \etal 1999a). 
It was one of the brightest GRB ever observed, ranking in the 0.3\% top of the BATSE flux distribution. Its distance
(z=1.6, Kulkarni \etal 1999) would 
imply a total energy of $1.6 \times10^{54}
erg$, assuming isotropical emission. This corresponds to $\sim 2 \Ms c^2$,
at the limit of all models of mergers (M\'esz\'aros  \& Rees   1998a).

This piece of evidence is lending support to the idea that, at least in
some case, the emission is collimated. This would reduce the energy
budget by $\sim \theta^2/4\pi$, where $\theta$ is the angle of the jet. 
A typical feature of a jet expansion (vs spherical) would be the presence
of an achromatic (i.e. energy-independent) break in the light curve,
 that appears when the relativistic beaming
angle $1/\Gamma$ becomes $\approx\theta$ (e.g. Rhoads 1997, Sari \etal 1999).
The presence of such a break has been claimed in 
GRB990123 (Kulkarni \etal 1999) and in
another more recent GRB, GRB990510 (Harrison \etal 1999).
With an angle $\theta\approx 10^{\circ}$, the total energy would
be reduced by $\approx 10^3$,   within
the limits of current models.
So far evidence of an achromatic break is limited 
to the optical range and an independent 
measurement confirming its presence in 
different regions of the spectrum
is lacking or not conclusive (e.g. in X-rays Kuulkers \etal 1999).

Very important indications on the geometry of the expansion
can be derived by comparing the
 prediction of the standard scenario (i.e. fireball expansion with
synchrotron emission) on the spectral and temporal
behaviour of the afterglow with observations.  
In particular, the spectral and temporal slopes of the afterglow emission 
$ F\sim t^{-\delta} \nu^{-\beta}$
are linked together by a relationship
that depends on the geometry of the fireball expansion
(Sari \etal 1997, 1999).
In the assumption of an adiabatic spherical expansion in a constant
density medium we have
$\delta=3\beta/2$ and $\delta=3\beta/2-1/2$  below and above the
cooling frequency $\nu_c$ respectively.
In the case of a jet expansion the relations are 
$\delta=2\beta+1$ ($\nu<\nu_c$) and $\delta=2\beta$ (($\nu>\nu_c$).
These relationship are plotted in fig.6  along with
the measured slopes we have derived for a first sample of X-ray
afterglows.
The data refer to a sample of 11 afterglows, 
among the brightest 
pointed by BeppoSAX upto May 1999, observed from few hours to about 
2 days after the GRB (Stratta \etal 1999, Piro \etal 2000).  
The average property of the sample are fully consistent with a
spherical expansion and deviates substantially from a jet
expansion {\it in the first two days}.
We stress that our sample is not biased against steep spectral slopes (
$\beta>1.5$), because $\gtsima 90\%$ of the GRB detected by BeppoSAX
and followed on with a fast observation, do show an X-ray
afterglow.

The disagreement with the jet prediction does not imply that
the geometry is spherical, because deviations from the
emission pattern of a
spherical expansion are expected only when the beaming angle
of the relativistic emission $\Gamma^{-1}$
becomes comparable to the opening angle of the jet
 $\theta_0$. This happens at a time
$t_{jet}\approx 6.2 (E_{52}/n_1)^{1/3}(\theta_0/0.1)^{8/3} hr$, which should
then be $\gtsima 48\ hr$.
We then derive $\theta_0\gtsima12\deg(n_1/E_{52})^{1/8}$.
Hence collimation, if present, cannot be very high.

\begin{figure}
\centerline{\epsfig{file=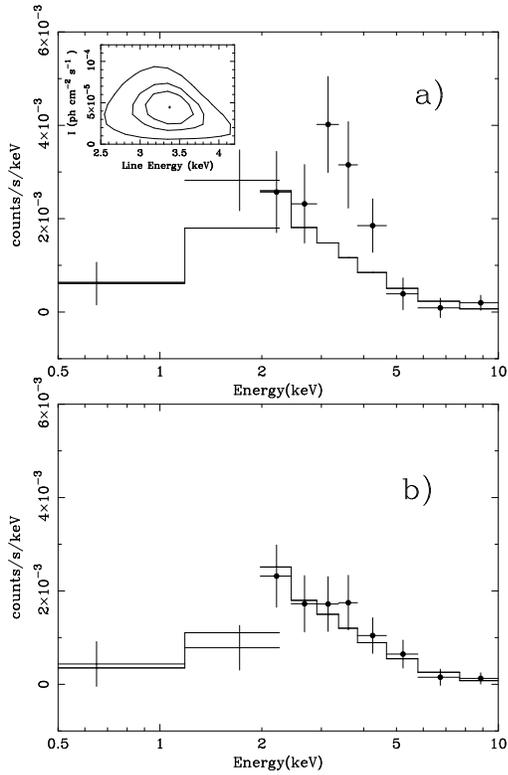,
width=9cm}}
\caption{Spectra (in detector counts) of the
X-ray afterglow of GRB970508 taken in the periods
0.2-0.6 day, (panel a) and 0.6-1 day (panel b) after the burst
The continuous line represents the best fit power law.
Note the feature around 3.5 keV in the first part (contour plots at 68\%, 90\%,
99\% in the inset) that, at a redshift of 0.83, corresponds
to the energy of the iron line complex (6.4-6.8 keV).
The line disappears in the second part of the observation (panel b)
(from Piro \etal 1999)}
\label{riga_fe}
\end{figure}

\section{The nature of the progenitor}
Information on the nature of the progenitor  can be drawn from
the GRB
environment. In the case of hypernova the massive star should die
young ($\approx 10^6\ $ years) and therefore GRB should be preferentially
hosted in regions near the center of star-forming galaxies.
On the contrary, NS-NS coalescence happens on much longer time scales and
the kick velocity given to the system by two consecutive supernova explosions
should bring a substantial fraction of these systems away from the the
parent galaxy. So far the angular displacement of 5  optical counterparts
 indicates that GRB are located within their host galaxies
(Bloom  \etal 1999a), 
favoring the association with star-forming regions. We note also
that those events are not located in the very center of their galaxies,
that excludes an association of GRB with AGN activity.

The other diagnostics of the  progenitor is based on spectral measurements
of broad and narrow features imprinted by a dusty - gas rich environment
expected in the hypernova scenario (e.g. Perna  \& Loeb  1998,
 M\'esz\'aros  \& Rees  1998b, Bottcher \etal 1998).
The absence of an optical transient in about 50\% of well localized
BeppoSAX GRB (25 as of Dec. 99), in which instead an X-ray afterglow
has been found in almost all the cases, may be explained by
heavy absorption by dust in the optical range, which leaves 
almost unaffected the X-rays (Owens  \etal 1998).

An exciting possibility is opened by the possible detection of 
X-ray iron line features in two different GRB, one by BeppoSAX (GRB970508
Piro \etal 1999b; Fig. 7) and the other
 by ASCA (GRB970828, Yoshida \etal 1999 ),
associated with  rebursting
on time scales of the order of a day.
It should be remarked that the presence of rebursting appears to be an
uncommon feature of X-ray afterglows, whose temporal behaviour
is very well described by power laws (Fig.8) at least
until 2-3 days, when the X-ray flux goes below the sensitivity
of current X-ray instruments.
Both the temporal and spectral features betray the presence of
dense ($n \sim 10^{10} cm^{-3}$) medium of $\approx 1 \Ms$
 near the site of the explosion ($\approx 10^{16} cm$) (Piro \etal
1999b). Such a medium should
have been pre-ejected before the GRB explosion, but the large
value of the density excludes stellar winds. A possible, intriguing explanation
is that the shell is the result of a SN explosion preceding the
GRB (Piro \etal 1999b, Vietri \etal 1999, Vietri \& Stella 1998).

\begin{figure}
\centerline{\epsfig{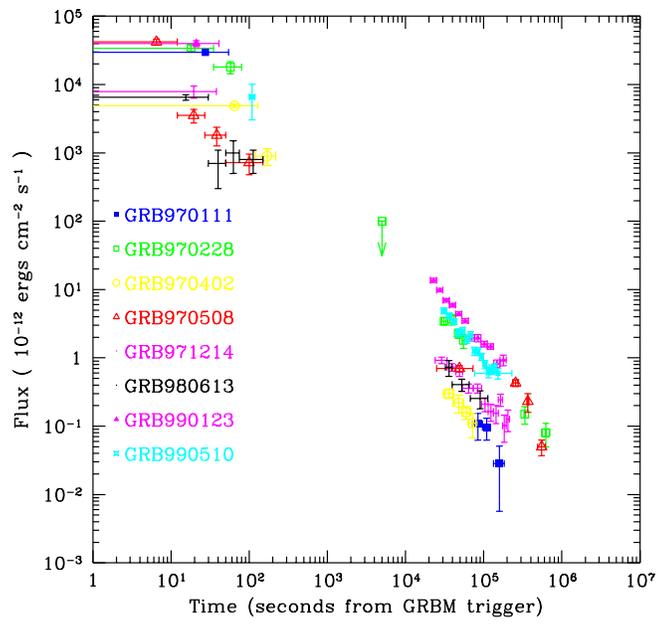}}
\caption{X-ray (2-10 keV) light curves of a sample of afterglows
by BeppoSAX
}
\label{aftall}
\end{figure}

Other evidences argue in favour of the association GRB-SN.
In the BeppoSAX error box of GRB980425 (Pian \etal 1999)
two groups (Kulkarni \etal 1998b, Galama \etal 1998) found a supernova
(SN1998bw) that had exploded at about the same time of the GRB. The probability of a chance coincidence of the two events is $\approx 10^{-4}$.
Since the majority of GRB are not associated with SN (e.g. Graziani \etal 1999), this
event (if the association is true) should apparently represent an uncommon 
kind of
GRB. However it is also possible that the two families are
indeed associated: this scenario would
require that the GRB are emitted by collimated jets.
The majority  of  GRB and  afterglow we see
are  beamed towards us, so that  
the contribution of the supernova to the total
emission is negligible.
The case of SN1998bw was then particular in that the jet producing the GRB
was collimated away from our line of sight, allowing the
detection of the (isotropic) SN emission at an early phase.
This scenario also explains  why GRB980425 was not particularly bright,
notwithstanding its redshift (z=0.0085), much lower than the typical
value of the other GRB ($z\approx 1$).
Since the afterglow decays as a power law, it is possible that at late
times the emission of the SN becomes detectable.
Evidence of such emission has been claimed in at least two cases
(GRB990326: Bloom \etal 1999b; GRB970228: Reichart \etal 1999)

\section{Conclusions}
Several bits of evidence supporting the
 association of GRB to star-forming regions
have been gathered so far. The potential perspectives
of this link are extremely exciting. Being  GRB 
the most powerful and distant sources of ionizing photons,
we can think of using them as probes of heavy elements
and star/galaxy formation in the early Universe.
A direct proof of this association is still missing but the
near future appears very promising in this respect.
BeppoSAX is discovering and localizing GRB and  X-ray afterglows
at a pace of 1 per month. Other satellites have also set up with
success procedures for rapid GRB localization (BATSE, XTE, ASCA
and IPN). The launch of HETE2, foreseen in early 2000
will increase substantially the number of well localized GRB.
Furthermore present and near-future big X-ray satellites, like Chandra, XMM,
ASTRO-E will allow detailed spectral studies of X-ray afterglows
and provide (Chandra) arcsec position of  X-ray counterparts
and, possibly, a direct redshift determination.

\begin{acknowledgements}

The  BeppoSAX results presented here were obtained through the joint
effort of all the components of the BeppoSAX Team.  
BeppoSAX is a major program of the Italian space agency (ASI) with
participation of the Netherlands agency for aerospace programs (NIVR)

\end{acknowledgements}

%

\end{document}